\documentclass[a4paper,twocolumn]{article}

\global\arraycolsep=2pt %reduces the separation in eqnarrays
\input{epsf}
\usepackage{graphicx}
\usepackage{cite}
\begin{document}
%-----------------------------
\makeatletter
\def\fmslash{\@ifnextchar[{\fmsl@sh}{\fmsl@sh[0mu]}}
\def\fmsl@sh[#1]#2{%
  \mathchoice
    {\@fmsl@sh\displaystyle{#1}{#2}}%
    {\@fmsl@sh\textstyle{#1}{#2}}%
    {\@fmsl@sh\scriptstyle{#1}{#2}}%
    {\@fmsl@sh\scriptscriptstyle{#1}{#2}}}
\def\@fmsl@sh#1#2#3{\m@th\ooalign{$\hfil#1\mkern#2/\hfil$\crcr$#1#3$}}
\makeatother
%--------------------------------
%---------------- CERN Titlepage <---------------------------
\thispagestyle{empty}
\begin{titlepage}

\vspace{0.3cm}
\boldmath
\begin{center}
  \Large {\bf Symmetry Breaking  \\ and  \\ Time Variation of the QCD Coupling}
\end{center}
\unboldmath
\vspace{0.8cm}

\unboldmath
\vspace{0.8cm}
\begin{center}
  {\large Harald Fritzsch}\\
 \end{center}
\begin{center}
{\sl Ludwig-Maximilians-University Munich, Sektion Physik}\\
{\sl Theresienstra{\ss}e 37, D-80333 Munich, Germany}\\
\end{center}
\vspace{\fill}
\begin{abstract}
\noindent
Astrophysical indications that the fine structure constant has
undergone a small time variation during the cosmological evolution are
discussed within the framework of the standard model of the
electroweak and strong interactions and of grand unification.  A
variation of the electromagnetic coupling constant could either be
generated by a corresponding time variation of the unified coupling
constant or by a time variation of the unification scale, of by both.
The various possibilities, differing substantially in their
implications for the variation of low energy physics parameters like
the nuclear mass scale, are discussed.  The case in which the
variation is caused by a time variation of the unification scale is of
special interest. It is supported in addition by recent hints towards
a time change of the proton-electron mass ratio.  
\end{abstract}
\end{titlepage}

The Standard Model of the electroweak and strong interactions has at
least 18 parameters, which have to be adjusted in accordance with
experimental observations. These include the three electroweak
coupling strengths $g_1$, $g_2$ $g_3$, the scale of the electroweak
symmetry breaking, given by the universal Fermi constant, the 9 Yukawa
couplings of the six quarks and the three charged leptons, and the
four electroweak mixing parameters.  One parameter, the mass of the
hypothetical scalar boson, is still undetermined. For the physics of
stable matter, i.e.  atomic physics, solid state physics and a large
part of nuclear physics, only six constants are of importance: the
mass of the electron, setting the scale of the Rydberg constant, the
masses of the $u$ and $d$-quarks setting the scale of the breaking of
isotopic spin, and the strong interaction coupling constant $\alpha_s$.
The latter, often parametrized by the QCD scale parameter $\Lambda$,
sets the scale for the nucleon mass. The mass of the strange quark can
also be included since the mass term of the $s$-quarks is expected to
contribute to the nucleon mass, although the exact amount of
strangeness contribution to the nucleon mass is still being discussed
- it can range from several tenth of MeV till more than 100 MeV.  As
far as macro-physical aspects are concerned, Newton's constant must be
added, which sets the scale for the Planck units of energy, space and
time.

Since within the Standard Model the number of free parameters cannot
be reduced, and thus far theoretical speculations about theories beyond
the model have not led to a well-defined framework, in view of lack of
guidance by experiment, one may consider the possibility that these
parameters are time and possibly also space variant on a cosmological
scale. Speculations about a time-change of coupling constants have a
long history, starting with early speculations about a cosmological
time change of Newton's constant $G$ \cite{Dirac,Milne,Jordan,Landau}.
Since in particular the masses of the fermions as well as the
electroweak mass scale are related to the vacuum expectation values of
a scalar field, time changes of these parameters are conceivable.  In
some theories beyond the Standard Model also the gauge coupling
constants are related to expectation values of scalar fields which
could be time dependent \cite{Damour:1994zq}.

Recent observations in astrophysics concerning the atomic
fine-structure of elements in distant objects suggest a time change of
the fine structure constant \cite{Webb:2001mn}.  The data suggest that
$\alpha$ was lower in the past, at a redshift of $z \approx 0.5 \ldots
3.5$:
\begin{equation} \label{exinput}
\Delta \alpha / \alpha = (-0.72 \pm 0.18) \times 10^{-5}.
\end{equation}

If $\alpha$ is indeed time dependent, the other two gauge coupling
constants of the Standard Model are also expected to depend on time,
as pointed out recently \cite{Calmet:2001nu} (see also
\cite{Dent:2001ga,Fritzsch:2002vz}), if the Standard Model is embedded
into a grand unified theory. Moreover the idea of a grand unification
of the coupling constants leads to a relation between the time
variation of the electromagnetic coupling constant and the QCD scale
parameter $\Lambda$, implying a physical time variation of the nucleon
mass, when measured in units given by an energy scale independent of
QCD, like the electron mass or the Planck mass.  The main assumption
is that the physics responsible for a cosmic time evolution of the
coupling constants takes place at energies above the unification
scale. This allows to use the usual relations from grand unified
theories to evolve the unified coupling constant down to low energy.
Of particular interest is the relatively large time change of the
proton mass in comparison to the time change of $\alpha$:
\begin{eqnarray} \label{result2}
\frac{\dot{M}}{M} \approx \frac{\dot{\Lambda}}{\Lambda}  \approx 38 \cdot \frac{\dot{\alpha}}{\alpha}.
\end{eqnarray}

Considering the six basic parameters mentioned above plus Newton's
constant G, one can in general consider seven relative time changes:
$\dot{G}/G$, $\dot{\alpha}/\alpha$, $\dot{\Lambda}/\Lambda$, $
\dot{m}_e / m_e$, $\dot{m}_u / m_u$, $\dot{m}_d / m_d$ and $\dot{m}_s
/ m_s$. Thus in principle seven different functions of time do enter
the discussion. However not all of them could be measured, even not in
principle.  Only dimensionless ratios e.g. the ratio $\Lambda/m_e$ or
the fine-structure constant could be considered as candidates for a
time variation.

The time derivative of the ratio $\Lambda/m_e$ describes a possible
time change of the atomic scale in comparison to the nuclear scale.
In the absence of quark masses there is only one mass scale in QCD,
unlike in atomic physics, where the two parameters $\alpha$ and $m_e$
enter.  The parameter $\alpha$ is directly measurable by comparing the
energy differences describing the atomic fine structure (of order $m_e
c^2 \alpha^4$) to the Rydberg energy $h c R_\infty=m_e c^2
\alpha^2/2\approx 13.606$ eV.

Both astrophysics experiments as well as high precision experiments in
atomic physics in the laboratory could in the future give indications
about a time variation of three dimensionless quantities: $\alpha$,
$M_p / m_e$ and $(M_n - M_p) / m_e$.
%If a time variation of all three
%ratios is seen, either the QCD scale $\Lambda$ changes with respect to
%the electron mass, or the quarks masses change with respect to it.
%Only the rate of change of the first ratio is much larger as the
%second one, one could conclude safely that $\Lambda$ is changing as
%well.
The time variation of $\alpha$ reported in \cite{Webb:2001mn} implies,
assuming a simple linear extrapolation, a relative rate of change per
year of about $1.0 \times 10^{-15}/$yr.  This poses a problem with
respect to the limit given by an analysis of the remains of the
naturally occurring nuclear reactor at Oklo in Gabon (Africa), which
was active close to 2 billion years ago.  One finds a limit of
$\dot{\alpha}/\alpha = (-0.2 \pm 0.8 ) \times 10^{-17} / \mbox{yr}$.
This limit was derived in \cite{Damour:1996zw} under the assumption
that other parameters, especially those related to the nuclear
physics, did not change during the last 2 billion years. It was
recently pointed out \cite{Calmet:2001nu,FS}, that this limit must be
reconsidered if a time change of nuclear physics parameters is taken
into account.  In particular it could be that the effects of a time
change of $\alpha$ are compensated by a time change of the nuclear
scale parameter.  For this reason we study in this paper several
scenarios for time changes of the QCD scale, depending on different
assumptions about the primary origin of the time variation.

Without a specific theoretical framework for the physics beyond the
Standard Model the relative time changes of the three dimensionless
numbers mentioned above are unrelated.  We shall incorporate the idea
of grand unification and assume for simplicity the simplest model of
this kind, consistent with present observations, the minimal extension
of the supersymmetric version of the Standard Model (MSSM), based on
the gauge group $SU(5)$.  In this model the three coupling constants
of the Standard Model converge at high energies at the scale
$\Lambda_G$.  In particular the QCD scale $\Lambda$ and the fine
structure constant $\alpha$ are related to each other. In the model
there are besides the electron mass and the quark masses three further
scales entering, the scale for the breaking of the electroweak
symmetry $\Lambda_w$, the scale of the onset of supersymmetry
$\Lambda_s$ and the scale $\Lambda_G$ where the grand unification sets
in. 

According to the renormalization group equations, considered here in
lowest order, the behaviour of the coupling constants changes
according to 
\begin{eqnarray} \label{running}
  \alpha_i(\mu)^{-1}
 &=&
  \left ( \frac{1}{\alpha^0_i(\Lambda_{G})}+\frac{1}{2 \pi}
  b^{S}_i  \ln
  \left ( \frac{\Lambda_{G}}{\mu} \right) \right) \theta(\mu- \Lambda_{S})
\\ \nonumber &&  +
\left ( \frac{1}{\alpha^0_i(\Lambda_{S})}+\frac{1}{2 \pi}
  b^{SM}_i   \ln
  \left ( \frac{\Lambda_{S}}{\mu} \right) \right)
\theta(\Lambda_{S}-\mu),
\end{eqnarray}
where the parameters $b_i$ are given by $b^{{SM}}_i\!\!\!=(b^{{SM}}_1,
b^{{SM}}_2, b^{{SM}}_3)=(41/10, -19/6, -7)$ below the supersymmetric
scale and by $b^{{S}}_i\!\!\!=(b^{{S}}_1, b^{{S}}_2, b^{{S}}_3)=(33/5,
1, -3)$ when ${\cal N}=1$ supersymmetry is restored, and where
\begin{eqnarray}
  \frac{1}{\alpha^0_i(\Lambda_{S})} &=&
\frac{1}{\alpha^0_i(M_Z)}
  +\frac{1}{2 \pi}
  b^{SM}_i \ln
  \left ( \frac{M_Z}{\Lambda_{S}} \right)
\end{eqnarray}
where $M_Z$ is the $Z$-boson mass and $\alpha^0_i(M_Z)$ is the value
of the coupling constant under consideration measured at $M_Z$.  We
use the following definitions for the coupling constants:
\begin{eqnarray} \label{def1}
  \alpha_1&=& 5/3 g_1^2/(4 \pi)=
  5 \alpha / ( 3 \cos^2(\theta)_{\overline{\mbox{MS}}})
    \\ \nonumber 
  \alpha_2&=& g_2^2/(4 \pi)= \alpha / \sin^2(\theta)_{\overline{\mbox{MS}}}
\\ \nonumber
\alpha_s&=& g_3^2/(4 \pi).
\end{eqnarray}

Assuming $\alpha_{u} = \alpha_u (t)$ and $\Lambda_G=\Lambda_G(t)$, one
finds:
\begin{eqnarray}
  \frac{1}{\alpha_i} \frac{\dot{\alpha}_i}{\alpha_i}=
  \left[\frac{1}{\alpha_u} \frac{\dot{\alpha}_u}{\alpha_u} 
 - \frac{b_i^S}{2 \pi} \frac{\dot{\Lambda}_G}{\Lambda_G}
  \right]
  \end{eqnarray}
which leads to
\begin{eqnarray}
  \frac{1}{\alpha} \frac{\dot{\alpha}}{\alpha}=
\frac{8}{3} \frac{1}{\alpha_s} \frac{\dot{\alpha}_s}{\alpha_s}
-\frac{1}{2 \pi} \left(b_2^S+\frac{5}{3} b_1^S-\frac{8}{3} b_3^S\right)
\frac{\dot{\Lambda}_G}{\Lambda_G}.
\end{eqnarray}
One may consider the following  scenarios:

\begin{enumerate}
\item[1)] $\Lambda_G$ invariant, $\alpha_u =\alpha_u (t)$. This is the
  case considered in \cite{Calmet:2001nu} (see also
  \cite{Dent:2001ga}), and one finds
  \begin{eqnarray}  \label{eq8}
  \frac{1}{\alpha} \frac{\dot{\alpha}}{\alpha}=
\frac{8}{3} \frac{1}{\alpha_s} \frac{\dot{\alpha}_s}{\alpha_s}
\end{eqnarray}
and
 \begin{eqnarray}
  \frac{\dot{\Lambda}}{\Lambda}= -\frac{3}{8} \frac{2 \pi}{b_3^{SM}}
  \frac{1}{\alpha}
    \frac{\dot{\alpha}}{\alpha}.
   \end{eqnarray}
\item[2)] $\alpha_u$ invariant, $\Lambda_G =\Lambda_G (t)$. One finds
  \begin{eqnarray} \label{eq10}
   \frac{1}{\alpha} \frac{\dot{\alpha}}{\alpha}=
-\frac{1}{2 \pi} \left(b_2^S+\frac{5}{3} b_1^S\right)
\frac{\dot{\Lambda}_G}{\Lambda_G}, 
\end{eqnarray}
with
\begin{eqnarray}
  \Lambda_G=\Lambda_S \left[\frac{\Lambda}{\Lambda_S} \mbox{exp}
    \left(-\frac{2 \pi}{b_3^{SM}} \frac{1}{\alpha_u} \right) \right]^{\left( \frac{b_3^{SM}}{b_3^{S}}\right)}
  \end{eqnarray}
  which follows from the extraction of the Landau pole using
  (\ref{running}). One obtains
  \begin{eqnarray} \label{eq12}
  \frac{\dot{\Lambda}}{\Lambda}= \frac{b_3^{S}}{b_3^{SM}}
  \left[ \frac{-2 \pi}{b_2^S+\frac{5}{3} b_1^S}\right]
  \frac{1}{\alpha} \frac{\dot{\alpha}}{\alpha} \approx -30.8
  \frac{\dot{\alpha}}{\alpha}
    \end{eqnarray}
  \item[3)] $\alpha_u =\alpha_u (t)$ and $\Lambda_G =\Lambda_G (t)$.
    One has
\begin{eqnarray}
    \frac{\dot{\Lambda}}{\Lambda}&=&
     -\frac{2 \pi}{b_3^{SM}} \frac{1}{\alpha_u}\frac{\dot{\alpha_u}}{\alpha_u}
     +\frac{b_3^{S}}{b_3^{SM}} \frac{\dot{\Lambda}_G}{\Lambda_G}
    \\ \nonumber &=& -\frac{3}{8} \frac{2
      \pi}{b_3^{SM}} \frac{1}{\alpha}\frac{\dot{\alpha}}{\alpha}
    -\frac{3}{8} \frac{1}{b_3^{SM}}
    \left(b_2^S+\frac{5}{3} b_1^S -\frac{8}{3} b_3^S \right)
    \frac{\dot{\Lambda}_G}{\Lambda_G}
    \\ \nonumber &=& 
46 \frac{\dot{\alpha}}{\alpha} + 1.07 \frac{\dot{\Lambda}_G}{\Lambda_G}
 \end{eqnarray}
 where theoretical uncertainties in the factor
 $R=(\dot{\Lambda}/\Lambda)/(\dot{\alpha}/\alpha)=46$ have been
 discussed in \cite{Calmet:2001nu}. The actual value of this factor is
 sensitive to the inclusion of the quark masses and the associated
 thresholds, just like in the determination of $\Lambda$. Furthermore
 higher order terms in the QCD evolution of $\alpha_s$ will play a
 role. In ref. \cite{Calmet:2001nu} it was estimated: $R=38\pm 6$.
 \end{enumerate}
 
  The case in which the time variation of $\alpha$ is not related to a
  time variation of the unified coupling constant, but rather to a
  time variation of the unification scale, is of particular interest.
  Unified theories, in which the Standard Model arises as a low energy
  approximation, might well provide a numerical value for the unified
  coupling constant, but allow for a smooth time variation of the
  unification scale, related in specific models to vacuum expectation
  values of scalar fields. Since the universe expands, one might
  expect a decrease of the unification scale due to a dilution of the
  scalar field. A lowering of $\Lambda_G$ implies according to
  (\ref{eq10})
  \begin{eqnarray} \label{eq19}
   \frac{\dot{\alpha}}{\alpha}= - \frac{1}{2 \pi}
\alpha \left(b_2^S+\frac{5}{3} b_1^S \right)
\frac{\dot{\Lambda}_G}{\Lambda_G}= -0.014 \frac{\dot{\Lambda}_G}{\Lambda_G}. 
\end{eqnarray}
If $\dot{\Lambda}_G / \Lambda_G$ is negative, $\dot{\alpha} / \alpha$
increases in time, consistent with the experimental observation.
Taking $ \Delta{\alpha} / \alpha=-0.72 \times 10^{-5}$, we would
conclude $\Delta{\Lambda}_G / \Lambda_G=5.1 \times 10^{-4}$, i.e. the
scale of grand unification about 8 billion years ago was about $8.3
\times 10^{12}$ GeV higher than today. If the rate of change is
extrapolated linearly, $\Lambda_G$ is decreasing at a rate
$\frac{\dot{\Lambda}_G}{\Lambda_G}=-7\times 10^{-14}/$yr.
%The rate for
%proton decay depends on $\Lambda_G$ as well as on $\Lambda$. The
%proton decay rate scales according to \cite{Ross:ai}
%\begin{eqnarray}
%  \Gamma_p= 4 \left( \frac{m_c m_s M_p^2}{2 v_1 v_2 \Lambda_G} \right )^2
%    b_0^2 A^2 \sin^4\theta (1.39 \times 10^{27}) \mbox{y}^{-1}
%    \propto \frac{M_p^4}{\Lambda^2_G} A^2
% \end{eqnarray}
% where $A$ is given by
%\begin{eqnarray}
%A&=& \left( \frac{\alpha_3(1\mbox{GeV})}{\alpha_3(m_c)} \right)^{2/7}
%   \left( \frac{\alpha_3(m_c)}{\alpha_3(m_b)} \right)^{6/25}
%   \left( \frac{\alpha_3(m_b)}{\alpha_3(m_t)} \right)^{6/23}
%   \left( \frac{\alpha_3(m_t)}{\alpha_3(m_W)} \right)^{2/7} \\ \nonumber &&
%   \left( \frac{\alpha_3(m_W)}{\alpha_u} \right)^{4/3}
%   \left( \frac{\alpha_2(m_W)}{\alpha_u} \right)^{-3}
%   \left( \frac{\alpha_1(m_W)}{\alpha_u} \right)^{-1/66}.
%\end{eqnarray}
%  The lifetime is expected to scale like $\dot \tau_p/\tau_p
% = -11.6 \frac{\dot{\alpha}}{\alpha}$.

 According to (\ref{eq12}) the relative changes of $\Lambda$ and
 $\alpha$ are opposite in sign. While $\alpha$ is increasing with a
 rate of $1.0 \times 10^{-15}/$yr, $\Lambda$ and the nucleon mass is
 decreasing, $\Lambda$ and the nucleon mass are decreasing, e.g. with
 a rate of $1.9 \times 10^{-14}/$yr. The magnetic moments of the
 proton $\mu_p$ as well of nuclei would increase according to
\begin{eqnarray}
\frac{\dot{\mu}_p}{\mu_p} = 30.8 \frac{\dot{\alpha}}{\alpha} \approx
3.1 \times 10^{-14}/ {\mbox yr}.
\end{eqnarray}
       
The effect can be seen by monitoring the ratio $\mu=M_p/m_e$.
Measuring the vibrational lines of H$_2$, a small effect was seen
\cite{Ivanchik:2001ji} recently. The data allow two different
interpretations:
\begin{itemize}
\item[a)] $\Delta \mu / \mu = (5.7 \pm 3.8)  \times 10^{-5}$
\item[b)] $\Delta \mu / \mu = (12.5 \pm 4.5) \times 10^{-5}$.
\end{itemize}
The interpretation b) agrees with the expectation based on
(\ref{eq12}):
\begin{eqnarray}
\frac{\Delta \mu}{\mu} =22 \times 10^{-5}.
\end{eqnarray}
It is interesting that the data suggest that $\mu$ is indeed
decreasing, while $\alpha$ seems to increase. If confirmed, this would
be a strong indication that the time variation of $\alpha$ at low
energies is caused by a time variation of the unification scale.

The time variation of the ratio $M_p/m_e$ and $\alpha$ discussed here
are such that they could by discovered by precise measurements in
quantum optics. The wave length of the light emitted in hyperfine
transitions, e.g. the ones used in the cesium clocks being
proportional to $\alpha^4 m_e/\Lambda$ will vary in time like
\begin{eqnarray}
\frac{\dot{\lambda}_{hf} }{\lambda_{hf}} = 4 \frac{\dot \alpha}{\alpha}
-\frac{\dot \Lambda}{\Lambda}\approx 3.5 \times 10^{-14}/\mbox{yr}
\end{eqnarray}
taking $\dot{\alpha}/\alpha\approx 1.0 \times 10^{-15}/$yr
\cite{Webb:2001mn}. The wavelength of the light emitted in atomic
transitions varies like $\alpha^{-2}$:
\begin{eqnarray}
\frac{\dot{\lambda}_{at} }{\lambda_{at}} = -2 \frac{\dot{\alpha} }{\alpha}.
\end{eqnarray}
One has ${\dot{\lambda}_{at} }/{\lambda_{at}}\approx
-2.0\times 10^{-15}/$yr. A comparison gives:
\begin{eqnarray}
  \frac{\dot{\lambda}_{hf}/\lambda_{hf}}{\dot{\lambda}_{at}/\lambda_{at}} = 
  -\frac{ 4 \dot{\alpha}/ \alpha - \dot \Lambda / \Lambda}{2 \dot{\alpha}/ \alpha }
  \approx -17.4.
\end{eqnarray}

%The wave length of the photons emitted by transitions between levels
%of molecular hydrogen will be time-dependent according to
%$\dot{\lambda}_{M}/\lambda_{M} \approx $ since here the scale of the
%levels is given by the atomic mass which is dominated by the proton
%mass.

At present the time unit second is defined as the duration of
6.192.631.770 cycles of microwave light emitted or absorbed by the
hyperfine transmission of cesium-133 atoms. If $\Lambda$ indeed
changes, as described in (\ref{eq12}), it would imply that the time
flow measured by the cesium clocks does not fully correspond with the
time flow defined by atomic transitions. %The effect, however, is small
%and corresponds to about 3 cesium cycles per day.

It remains to be seen whether the effects discussed in this paper
can soon be observed in astrophysics or in quantum optics. A
determination of the double ratio
$(\dot{\Lambda}/\Lambda)/(\dot{\alpha}/\alpha)=R$ would be of crucial
importance, both in sign and in magnitude. If one finds the ratio to
be about $- 20$, it would be  considered as a strong indication of a
unification of the strong and electroweak interactions based on a
supersymmetric extension of the Standard Model. In any case the
numerical value of $R$ would be of high interest towards a better
theoretical understanding of time variation and unification.
\section*{Acknowledgements}
We should like to thank A. Albrecht, J. D. Bjorken, E. Bloom, G.
Boerner, S. Brodsky, P. Chen, S.  Drell, G. Goldhaber, T. Haensch, M.
Jacob, P. Minkowski, A.  Odian, H. Walther and A. Wolfe for useful
discussions.


\begin{thebibliography}{10}  
%%%%%%%%%%%%%%%%%%%%%%%%%%%%%%%%%%%%%%%%%%%%%%%%%%%%%%%%%%%%%%%%%%%%%%%%%%%%%
\bibitem{Dirac} P.~M.~Dirac, Nature {\bf192}, 235 (1937).
\bibitem{Milne} E.~A.~Milne, Relativity, Gravitation and World
  Structure, Clarendon press, Oxford, (1935), Proc. Roy. Soc. A, 3,
  242 (1937).  \bibitem{Jordan} P.~Jordan, Naturwiss., 25, 513 (1937),
  Z. Physik, 113, 660 (1939).  \bibitem{Landau} L.~D.~Landau, in
  W.~Pauli, ed., Niels Bohr and the Development of Physics,
  McGraw-Hill, NY, p.52, (1955).
%%%%%%%%%%%%%%%%%%%%%%%%%%%%%%%%%%%%%%%%%%%%%%%%%%%%%%%%%%%%%%%%%%%%%%%%%%%%%

%%%%%%%%%%%%%
%\cite{Damour:1994zq}
\bibitem{Damour:1994zq}
  %\cite{Green:sp}
%\bibitem{Green:sp}
M.~B.~Green, J.~H.~Schwarz and E.~Witten, Superstring Theory, Vol. 1 and 2,
T.~Damour and A.~M.~Polyakov,
%``The String dilaton and a least coupling principle,''
Nucl.\ Phys.\ B {\bf 423}, 532 (1994)
[arXiv:hep-th/9401069].
%%CITATION = HEP-TH 9401069;%%
%%%%%%%%%%%%%
  
%%%%%%%%%%%%%%%%%%%%%%%%%%%%%%%%%%%%%%%%%%%%%%%%%%%%%%%%%%%%%%%%%%%%%% 
%\cite{Webb:2001mn}
\bibitem{Webb:2001mn}
J.~K.~Webb {\it et al.},
%``Further Evidence for Cosmological Evolution of the Fine Structure Constant,''
Phys.\ Rev.\ Lett.\  {\bf 87} 091301 (2001) 
[arXiv:astro-ph/0012539].
%%CITATION = ASTRO-PH 0012539;%%
%%%%%%%%%%%%%%%%%%%%%%%%%%%%%%%%%%%%%%%%%%%%%%%%%%%%%%%%%%%%%%%%%%%%%%%%%%%%%

%%%%%%%%%%%%%%%%%%%%%%%%%%%%%%%%%%%%%%%%%%%%%%%%%%%%%%%%%%%%%%%%%%%%%%%%%%%%%
%%%%%%%%%%%%%
%\cite{Calmet:2001nu}
\bibitem{Calmet:2001nu}
X.~Calmet and H.~Fritzsch,
%``The cosmological evolution of the nucleon mass and the electroweak  coupling
%constants,''
arXiv:hep-ph/0112110, to appear in Eur.Phys. J. C.
%%CITATION = HEP-PH 0112110;%%
%%%%%%%%%%%%%         
%%%%%%%%%%%%%%%%%%%%%%%%%%%%%%%%%%%%%%%%%%%%%%%%%%%%%%%%%%%%%%%%%%%%%%%%%%%%%


%%%%%%%%%%%%%%%%%%%%%%%%%%%%%%%%%%%%%%%%%%%%%%%%%%%%%%%%%%%%%%%%%%%%%%%%%%%%%
%%%%%%%%%%%%%
%\cite{Dent:2001ga}
\bibitem{Dent:2001ga}

%%%%%%%%%%%%%%%%%%%%%%%%%%%%%%%%%%%%%%%%%%%%%%%%%%%%%%%%%%%%%%%%%%%%%%%%%%%%%
%%%%%%%%%%%%%
%\cite{Langacker:2001td}
%\bibitem{Langacker:2001td}
P.~Langacker, G.~Segre and M.~J.~Strassler,
%``Implications of gauge unification for time variation of the fine  structure
%constant,''
Phys.\ Lett.\ B {\bf 528}, 121 (2002)
[arXiv:hep-ph/0112233];
%%CITATION = HEP-PH 0112233;%%
%%%%%%%%%%%%%           
T.~Dent and M.~Fairbairn,
%``Time-varying coupling strengths, nuclear forces and unification,''
arXiv:hep-ph/0112279.
%%CITATION = HEP-PH 0112279;%%
%%%%%%%%%%%%%      
%%%%%%%%%%%%%
%\cite{Fritzsch:2002vz}
\bibitem{Fritzsch:2002vz} H.~Fritzsch, ``Fundamental constants at high
energy,'' arXiv:hep-ph/0201198, Invited talk at Symposium on 100 Years
Werner Heisenberg: Works and Impact, Bamberg, Germany, 26-30 Sep 2001;
%%CITATION = HEP-PH 0201198;%%
%%%%%%%%%%%%%                              
%%%%%%%%%%%%%
%\cite{Landau:2000cc}
%\bibitem{Landau:2000cc}
S.~J.~Landau and H.~Vucetich,
%``Testing theories that predict time variation of fundamental constants,''
arXiv:astro-ph/0005316;
%%CITATION = ASTRO-PH 0005316;%%
%%%%%%%%%%%%%%%%%%%%%%%%%%
%\cite{Flambaum:2002de}
%\bibitem{Flambaum:2002de}
V.~V.~Flambaum and E.~V.~Shuryak,
%``Limits on cosmological variation of strong interaction and quark masses  from
%big bang nucleosynthesis, cosmic, laboratory and Oklo data,''
arXiv:hep-ph/0201303;
%%CITATION = HEP-PH 0201303;%%
%%%%%%%%%%%%%
%%%%%%%%%%%%%%%%%%%%%%%%%%%%%%%%%%%%%%%%%%%%%%%%%%%%%%%%%%%%%%%%%%%%%%%%%%%%%
%%%%%%%%%%%%%%%%%%%%%%%%%%%%%%%%%%%%%%%%%%%%%%%%%%%%%%%%%%%%%%%%%%%%%%%%%%%%%
%\cite{Wetterich:2002ic}
%\bibitem{Wetterich:2002ic}
C.~Wetterich,
%``Probing quintessence with time variation of couplings,''
arXiv:hep-ph/0203266;
%%CITATION = HEP-PH 0203266;%%
%\cite{Chacko:2002mf}
%\bibitem{Chacko:2002mf}
Z.~Chacko, C.~Grojean and M.~Perelstein,
%``Fine structure constant variation from a late phase transition,''
arXiv:hep-ph/0204142.
%%CITATION = HEP-PH 0204142;%%
%%%%%%%%%%%%%%%%%%%%%%%%%%%%%%%%%%%%%%%%%%%%%%%%%%%%%%%%%%%%%%%%%%%%%%%%%%%%%

%%%%%%%%%%%%%
\bibitem{FS}
V.~V.~Flambaum and E.~V.~Shuryak, in \cite{Fritzsch:2002vz}.
%%%%%%%%%%%%%

%%%%%%%%%%%%%%%%%%%%%%%%%%%%%%%%%%%%%%%%%%%%%%%%%%%%%%%%%%%%%%%%%%%%%%%%%%%%%
%\cite{Damour:1996zw}
\bibitem{Damour:1996zw}
T.~Damour and F.~Dyson,
%``The Oklo bound on the time variation of the fine-structure constant  revisited,''
Nucl.\ Phys.\ B {\bf 480}, 37 (1996)
[arXiv:hep-ph/9606486].
%%CITATION = HEP-PH 9606486;%%
%%%%%%%%%%%%%%%%%%%%%%%%%%%%%%%%%%%%%%%%%%%%%%%%%%%%%%%%%%%%%%%%%%%%%%%%%%%%%

%%%%%%%%%%%%%%%%%%%%%%%%%%%%%%%%%%%%%%%%%%%%%%%%%%%%%%%%%%%%%%%%%%%%%%%%%%%%%
%\cite{Ross:ai}
%\bibitem{Ross:ai}
%G.~G.~Ross,
%``Grand Unified Theories,''
%{\it  Reading, Usa: Benjamin/cummings ( 1984) 497 P. ( Frontiers In Physics, 60)}.
%%%%%%%%%%%%%%%%%%%%%%%%%%%%%%%%%%%%%%%%%%%%%%%%%%%%%%%%%%%%%%%%%%%%%%%%%%%%%


%%%%%%%%%%%%%%%%%%%%%%%%%%%%%%%%%%%%%%%%%%%%%%%%%%%%%%%%%%%%%%%%%%%%%%%%%%%%%
%\cite{Ivanchik:2001ji}
\bibitem{Ivanchik:2001ji} A.~V.~Ivanchik, E.~Rodriguez, P.~Petitjean
  and D.~A.~Varshalovich,
%``Do the fundamental constants vary in the course of the cosmological  evolution?,''
arXiv:astro-ph/0112323.
%%CITATION = ASTRO-PH 0112323;%%
%%%%%%%%%%%%%%%%%%%%%%%%%%%%%%%%%%%%%%%%%%%%%%%%%%%%%%%%%%%%%%%%%%%%%%%%%%%%%



\end{thebibliography}
\end{document}